\documentclass[aps,prl,amsmath,two column,amssymb,floatfixng,showpacs,
superscriptaddress,longbibliography,footinbib]{revtex4-1}
\pdfoutput=1
\usepackage[dvips]{graphics}
\usepackage{bm}
\usepackage{float}
\usepackage{epsfig}
\usepackage{enumerate}
\usepackage{amsmath}
\usepackage{color}
\usepackage{graphicx}
\usepackage[colorlinks=true,linktoc=page,linkcolor=red,citecolor=blue]{hyperref}
\usepackage{natbib}
\usepackage{float}
\usepackage{graphicx}
\usepackage{dcolumn}
\usepackage{bm}
\usepackage{latexsym,epsfig}
\usepackage{graphicx}
\usepackage{verbatim}
\usepackage{comment}
\usepackage{amsmath}
\usepackage[utf8]{inputenc}
\usepackage{amssymb}
\usepackage{physics}
\usepackage{stmaryrd}
\usepackage{color}
\usepackage{epstopdf}
\usepackage{grffile}
\usepackage{lipsum}
\usepackage{enumitem}
\usepackage[dvipsnames]{xcolor}
\DeclareGraphicsExtensions{.eps}

\newcommand{\beq}{\begin{equation}}
\newcommand{\eeq}{\end{equation}}
\newcommand{\bea}{\begin{eqnarray}}
\newcommand{\eea}{\end{eqnarray}}
\newcommand{\ben}{\begin{eqnarray*}}
\newcommand{\een}{\end{eqnarray*}}
\newcommand{\bfig}{\begin{figure}}
\newcommand{\efig}{\end{figure}}
\usepackage{soul}

\usepackage{hyperref}
\hypersetup{
    colorlinks=true,      
    urlcolor=blue,
    citecolor=blue,
    linkcolor=blue
}

\begin{document}

\title{Spin-aligned butterfly spectral map in Non-Hermitian quasicrystals }
\author{Soumya Ranjan Padhi}
\affiliation{School of Physical Sciences, National Institute of Science Education and Research, Jatni 752050, India}
\affiliation{Homi Bhabha National Institute, Training School Complex, Anushaktinagar, Mumbai 400094, India}
\author{Souvik Roy}
\affiliation{School of Physical Sciences, National Institute of Science Education and Research, Jatni 752050, India}
\affiliation{Homi Bhabha National Institute, Training School Complex, Anushaktinagar, Mumbai 400094, India}
\author{Debashree Chowdhury}
\email{debashreephys@gmail.com}
\affiliation{Centre for Nanotechnology, IIT Roorkee, Roorkee, Uttarakhand  247667, India }
\author{Tapan Mishra}
\email{mishratapan@gmail.com}
\affiliation{School of Physical Sciences, National Institute of Science Education and Research, Jatni 752050, India}
\affiliation{Homi Bhabha National Institute, Training School Complex, Anushaktinagar, Mumbai 400094, India}
\date{\today}

\begin{abstract}

The Non-Hermitian spinful Aubry-Andr\'e-Harper (AAH) model in the presence of Rashba-type spin-orbit coupling (RSOC) and a spatially varying textured magnetic field is studied. Interestingly, our analysis produces a butterfly spectral map due to the non-trivial extent of localization of the states in the spectrum. This spectral map also exhibits an asymmetric spin alignment with respect to the wings of the butterfly. Our analysis also suggests that the onset of such a spectral map is a combined effect of the non-hermiticity, spin-orbit interaction, and the textured magnetic field.

\end{abstract}

\maketitle

\paragraph*{Introduction.-} Non-Hermitian systems are at the forefront of research because of their non-standard physical properties and immense potential for technological application ~\cite{NHA, NHA1, NHA2}. Notable properties include the appearance of complex spectra~\cite{Bergholtz}, exceptional degeneracies~\cite{Bergholtz}, broken bulk-boundary correspondence~\cite{Lee2016AnomalousES, Shen2018, Takata2018}, skin effect~\cite{Yao_prl_2018, Yuce_2020, Masatoshi_prl_2020, Li_2020, Okuma_2023, Li_2021, Lin_2023, Paolo_2023, Robert_2020, Emil_2021, Andre_2019, Nori_2019, Alvarez_2018, Masahito_prx_2018, Emil_2018, Koch_2020, Cao_2021, Xiao_2020, Jin_2019} and comb effects~\cite{padhi1,padhi2}. 
When combined with disorder, non-Hermitian systems open new fertile ground to observe novel scenarios in the context of localization transitions, which are typically absent in their Hermitian counterparts. In this context, one-dimensional non-Hermitian quasiperiodic  (intermediate to random and periodic) lattices have been widely explored, revealing numerous interesting phenomena. The simplest among them is the one-dimensional tight-binding lattice with an onsite Aubry-Andr\'e (AA) type potential. When Hermitian, such a model exhibits a transition of the entire extended spectrum to localization after a critical quasiperiodic potential strength~\cite{Aubry1980Analyticity, Harper1955, Andre_2019,Paredesreview_2019, Roati2008}. However, when the AA potential is complex, in other words, when the model becomes non-Hermitian, it exhibits a delocalization-to-localization transition at a critical non-Hermiticity parameter, even though the potential strength is fixed~\cite{Longhi_prl_2019}. Such non-trivial behaviour sparked enormous interest in exploring non-Hermitian quasiperiodic lattices, especially in one-dimensional systems~\cite{Zhang_pra_2021, Cai_2022, Shuchen_es_2021, Zhou_prb_2023, Soutang_rent_2023, Zhou_dimerization_2022, Wu_iop_2021, Zeng_prb_2020, ShuChen_EMEs_2021, ShuChen_Exp_decay_2021, Wang_unconventional_2021, Longhi_maryland_2021, Zhou_floquet_2021, Zhou_marryland_2022, Zeng_GMEs_NHGAA_expt_2020, Li_mobrings_2024, Padhan_prbl_nh_2023, Gandhi_kitaev_2024, Peng_power_nh_2023, ShuChen_power_nh_2021, Tong_liu_2020, AMES_arxiv_2024, RSO_NH_1}, resulting in numerous novel theoretical findings in recent years, followed by several experimental observations in various quantum simulators~\cite{E1, E2, ghatak2024, Weidemann2022}. 

In this context, the system of ultracold atoms in optical lattices is proven to be a versatile platform to study localization transitions. Due to the flexibility in manipulating the lattice parameters and exquisite control over the laser-atom interactions, the localization transitions have been observed in different contexts, ranging from non-interacting to many-body systems. Adding to these findings, spinful electrons in quasiperiodic lattices have also attracted a great deal of attention in recent years. For example, in Refs. \cite{HS1,HS}, the AA model has been discussed in the presence of spin-orbit coupling (SOC). Ref.~\cite{HS},  mentions that in the presence of Rashba-type SOC, the AA model shows self-duality and metal-insulator transition is observed at a critical strength of the quasiperiodic potential. It is shown in Ref.~\cite{HS1} that in the absence of any external magnetic field, SOC brings a modified hopping amplitude; the scenario is changed in presence of a Zeeman field.  The interplay of SOC and Zeeman field leads to interesting effects on the localization length and Lyapunov exponent.  While the Hermitian quasiperiodic systems in spin-orbit coupled systems have been explored widely, the effect of non-Hermiticity in such systems has not been well explored. A recent experimental study \cite{Zhao2025Localization} demonstrated the emergence of distinct phases in a Raman lattice system for alkaline-earth-like atoms by examining the effects of both Hermitian and non-Hermitian conditions. The system incorporates incommensurate Zeeman potential and dissipation for spin up and down, enabling the examination of critical phases and mobility edges. 
This work motivates us to study a spin-full non-Hermitian AAH system in the presence of spin-orbit interaction and Zeeman field.

By considering Rashba-type spin-orbit coupling (RSOC) and a spatially varying textured magnetic field that rotates the electron spin along a ring, we study the localization transition by tuning the non-Hermitian quasiperiodic potential. We show that under proper condition, the system exhibits a delocalization to localization transition through an intermediate phase that hosts both localized and delocalized states. Interestingly, in the localized phase close to the transition boundary, we obtain a peculiar localization behaviour of the states in the spectrum which together map out a butterfly-type pattern for some particular values of the non-Hermitian potential. We call this novel behaviour the butterfly spectral map and this is the key focus of our analysis. We also show that the states exhibit an asymmetric spin alignment along the wings of the butterfly due to the textured magnetic field. Counterintuitively,  with an increase in the NH potential, the butterfly spectral map disappears into a rectangular map at a very strong NH potential, avoiding a strong localization of the entire spectrum. In a recent paper, it has been shown that the appearance of non-Hermitian butterfly spectra in quasiperiodic systems \cite{BFS}. The difference of their finding from ours lies in the fact that i) our butterfly spectra appear as a combined effect of non-hermiticity and a textured magnetic field,\\
ii) in our case, the butterfly spectra appear in the IPR distribution with the real part of the energy and not in the complex plane (between real and imaginary energies) \cite{BFS}.

\begin{figure}
    \centering
    \includegraphics[width=0.85\linewidth]{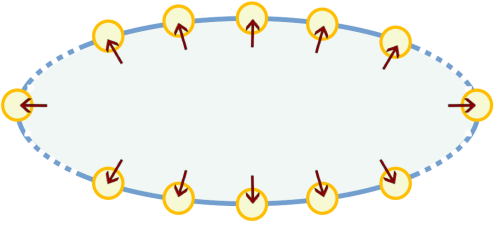}
    \caption{Schematic representation of the textured spin configuration, highlighting the spatial pattern and directional alignment of spins within the system.}
    \label{fig:1}
\end{figure}

\paragraph*{Model.-} We consider a one-dimensional tight-binding ring of $N$ sites with spin-$1/2$ fermions. The total Hamiltonian is written as
\begin{equation}
\mathcal{H} = \mathcal{H}_A + \mathcal{H_R} + \mathcal{H_\sigma},
\end{equation}
where $\mathcal{H}_A$ accounts for the quasiperiodic Aubry–Andr\'e potential, $\mathcal{H_R}$ introduces the Rashba spin–orbit coupling and $\mathcal{H_\sigma}$ incorporates the spatially varying spin orientation along the ring. The textured spin configuration and its directional alignment are indicated in Fig.~\ref{fig:1} schematically. 
The Aubry–Andr\'e contribution is
\begin{equation}
\mathcal{H}_A = -t \sum_{n=1}^{N} \sum_{\sigma} 
\left( c^\dagger_{n+1,\sigma} c_{n,\sigma} + \text{H.c.} \right) 
+ \sum_{n=1}^{N} \sum_{\sigma} V_n c^\dagger_{n,\sigma} c_{n,\sigma},
\end{equation}
with onsite potential
\begin{equation}
V_n = \lambda \cos(2\pi \alpha n + \phi),
\qquad \phi = \theta + i \Gamma,
\end{equation}
where $c_{n,\sigma}^\dagger$ ($c_{n,\sigma}$) creates (annihilates) a fermion with spin $\sigma=\uparrow,\downarrow$ at site $n$ and $\Gamma $ is the non-Hermiticity parameter. $\lambda$  is the quasi-periodic potential strength. Here $\alpha$ is the ratio of two consecutive Fibonacci numbers. We choose $\theta=0$ for our entire analysis.

The Rashba spin–orbit interaction is given by
\begin{align}
\mathcal{H}_R &= -\alpha_z \sum_{n=1}^{N} \sum_{\sigma,\sigma'}
\left[ c^\dagger_{n+1,\sigma}\, (i\sigma_y)_{\sigma\sigma'}\, c_{n,\sigma'} + \text{H.c.} \right] \nonumber\\
&\quad - \alpha_y \sum_{n=1}^{N} \sum_{\sigma,\sigma'}
\left[ c^\dagger_{n+1,\sigma}\, (i\sigma_z)_{\sigma\sigma'}\, c_{n,\sigma'} + \text{H.c.} \right].
\end{align}
where $\alpha_y$ and $\alpha_z$ are the Rashba coupling strengths, and $\sigma_y$, $\sigma_z$ are Pauli matrices. This term generates spin-dependent hopping and breaks spin conservation, mimicking Rashba-type spin–orbit coupling in a one-dimensional lattice.

Finally, the spin-texture Hamiltonian which is the Zeeman field, reads as,
\begin{equation}
\mathcal{H}_\sigma = \sum_{n=1}^{N}
c_{n}^\dagger \left[ \mathbf{h}_n \cdot \boldsymbol{\sigma} \right] c_{n},
\end{equation}
with spinor $c_{n}^\dagger = (c_{n,\uparrow}^\dagger, c_{n,\downarrow}^\dagger)$, Pauli vector $\boldsymbol{\sigma}=(\sigma_x,\sigma_y,\sigma_z)$, and site-dependent effective field
\begin{equation}
\mathbf{h}_n = h_z  \big( \sin \theta_n , 0 , \cos \theta_n \big),
\qquad \theta_n = \frac{2\pi n}{N}.
\end{equation}
This parametrization ensures that the spin points along $+\hat{z}$ at site $n=1$, gradually rotates through the $xz$-plane, and aligns along $-\hat{z}$ at site $n=N$, thereby introducing a controlled spin-flip profile across the system.

The full Hamiltonian describes a non-Hermitian Aubry–Andr\'e model enriched with Rashba spin–orbit coupling and a site-dependent spin texture. 
In the following, we show that combined action of the quasiperiodic potential, the spatially varying spin texture, and the spin-orbit-induced spin mixing offers a versatile platform for exploring localization–delocalization transitions, spin-flip processes, and the emergence of nontrivial spectral map.

\paragraph*{Results.-} We first present the scenario of localization transition in the system described by Eq.~(1) by investigating the eigenstates and eigenenergies derived from the exact solution of the Schrödinger equation, \( \mathcal{H}\psi_j = E_j\psi_j \). This equation governs the quantum behavior of the system, where $\mathcal{H}$  is the Hamiltonian, \( |\psi_j\rangle \) represents the j-th eigenstate, and E is the corresponding eigenenergy. The nature of each eigenstate, whether localized or delocalized, is determined using two key metrics, the  inverse participation ratio (IPR) and the normalized participation ratio (NPR).
The IPR is mathematically expressed as
\begin{equation} \text{IPR}_j = \frac{\sum_{n=1}^{N} |\psi_{j,n\uparrow}|^4 + |\psi_{j, n\downarrow}|^4}{(\langle \psi_{j}|\psi_{j} \rangle)^2}, \end{equation}
denotes the amplitude of the eigenstate at site. NPR is the normalized participation ratio (NPR), which is defined as
\begin{equation} \text{NPR}_j = (N \times \text{IPR}_j)^{-1}. \end{equation}

In the thermodynamic limit, the behavior of these measures distinguishes delocalized and localized eigenstates.
For delocalized eigenstates, the probability density spreads uniformly across the system and thus IPR $\approx0$ (NPR remains finite). For localized eigenstates, since the eigenstate remains confined to a small region IPR remains finite (NPR $\approx0$). To obtain a comprehensive understanding of the entire spectrum, we compute the average IPR and average NPR by summing these measures over all eigenstates. These are defined as,
\begin{equation} \langle \text{IPR} \rangle = \frac{1}{2N} \sum_{j=1}^{2N} \text{IPR}_{j}, \quad \langle \text{NPR} \rangle = \frac{1}{2N} \sum_{j=1}^{2N} \text{NPR}_j, \end{equation}
where $N$ is the total number of eigenstates in the system. The averaged quantities provide global insights into the nature of the spectrum, revealing whether the system predominantly exhibits localized, delocalized, or mixed-state behavior. 
The complete phase diagram is constructed by evaluating the parameter $\eta$, defined as 
\[
\eta = \log_{10} (\langle \mathrm{IPR} \rangle \times \langle \mathrm{NPR} \rangle).
\]

\begin{figure}[t]
    \centering
    \includegraphics[width=1.0\linewidth]{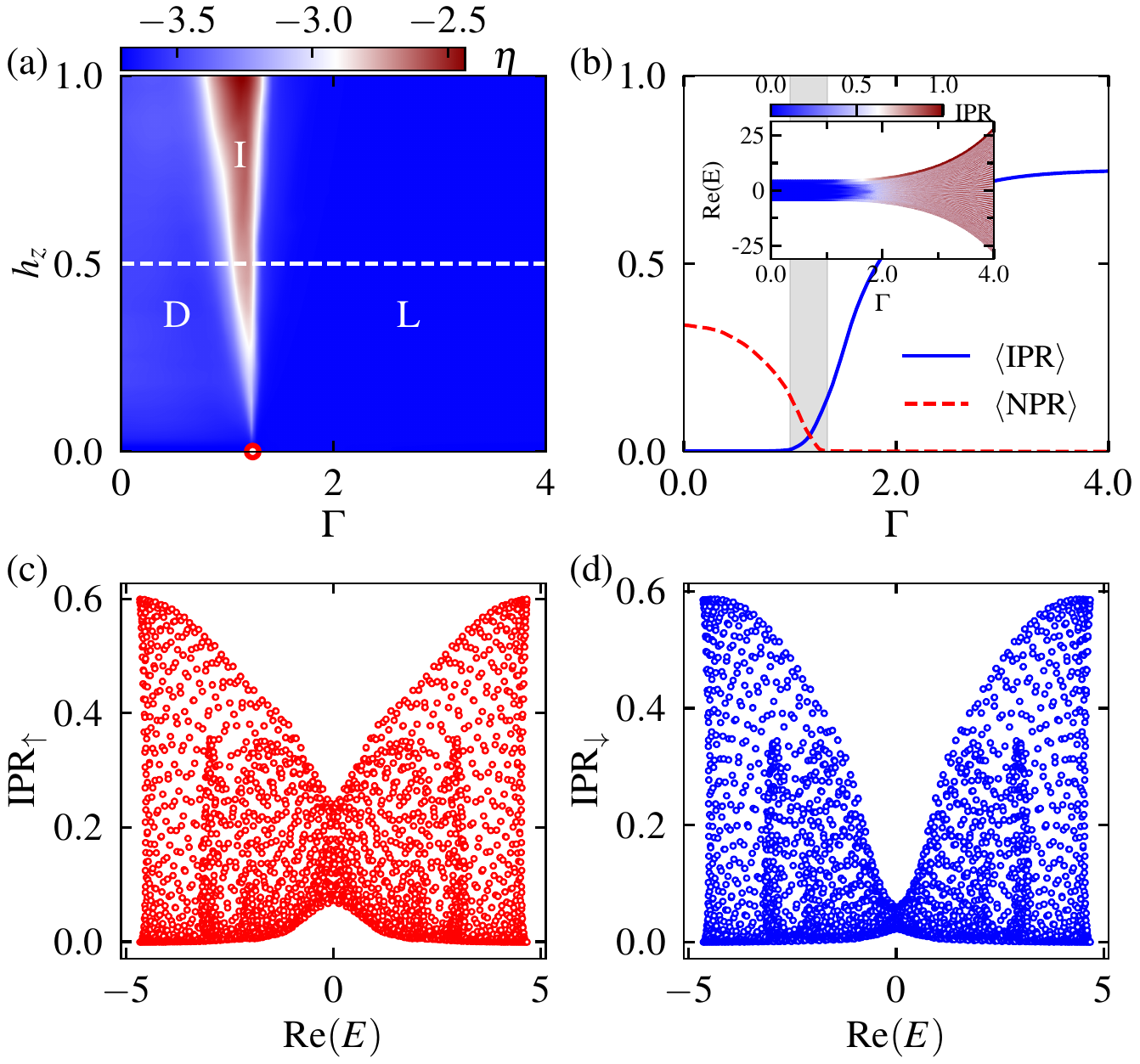}
    \caption{(a) Phase diagram with the corresponding $\eta$ values in the $h_z$–$\Gamma$ plane with $\alpha_y=\alpha_z=1.0$ indicating the delocalized (D), localized (L), and intermediate (I) region. The red circle indicates the critical point ($\Gamma \approx 1.24$). (b) $\langle \text{IPR} \rangle$ (blue solid line) and $\langle \text{NPR} \rangle$ (red dashed line) are plotted as a function of the complex phase $\Gamma$. The gray region is the intermediate phase. The inset shows the IPR of all the state as function of their real eigenenergies and  $\Gamma$ . (c) and (d) IPR of the two spin components up and down as a function of the real part of eigenenergies. The other parameters are fixed as $\lambda=1$ and system size $N=2584$.}
    \label{fig:2}
\end{figure}

It should be noted here that our system has three tuning parameters, namely the non-Hermitian parameter ($\Gamma$), the magnetic parameter ($h_{z}$), and the RSOC ($\alpha_{y,z}$). The RSOC strength is maintained at a finite level with $\alpha_y=\alpha_z=1$ to investigate the impact of both RSO and the Zeeman field $h_z$. 
In Fig. \ref{fig:2} (a), we depict the phase diagram in the $h_{z}- \Gamma$ plane along with the corresponding $\eta$ values of the spectrum. The intermediate region (I), characterized by the maximum $\eta$ values, is shown in red, while the delocalized (D) and localized (L) regions are indicated by the minimum value of $\eta$. The phase diagram depicts that when $h_z=0$, a sharp delocalization to localization transition occurs at $\Gamma\approx1.24$ (shown by a red circle in Fig. \ref{fig:2} (a)). However, for any finite values of $h_z$, an intermediate phase appears which becomes wider with an increase in $h_z$. To substantiate this, we plot $\langle \text{IPR} \rangle$ (blue solid line) and $\langle \text{NPR} \rangle$ (red dashed line) as functions of $\Gamma$ in Fig. \ref{fig:2} (b), for a cut through the phase diagram at $h_{z}=0.5$ (dashed line in Fig. \ref{fig:2} (a)). Clearly, the intermediate phase appears for $1.0 \lesssim \Gamma \lesssim 1.35$ separating the delocalized and localized phases on its left and right regions, respectively. The intermediate nature of the spectrum is shown in the inset of Fig. \ref{fig:2} (b), where the IPR of states plotted with respect to their real eigen energies and $\Gamma$ where for a range of $\Gamma$ the spectrum hosts both localized and delocalized states. This indicates the crucial role played by the Zeeman field in stabilising the intermediate phase together with the RSOC. 

\begin{figure*}[t]
    \centering
    \includegraphics[width=1.0\linewidth]{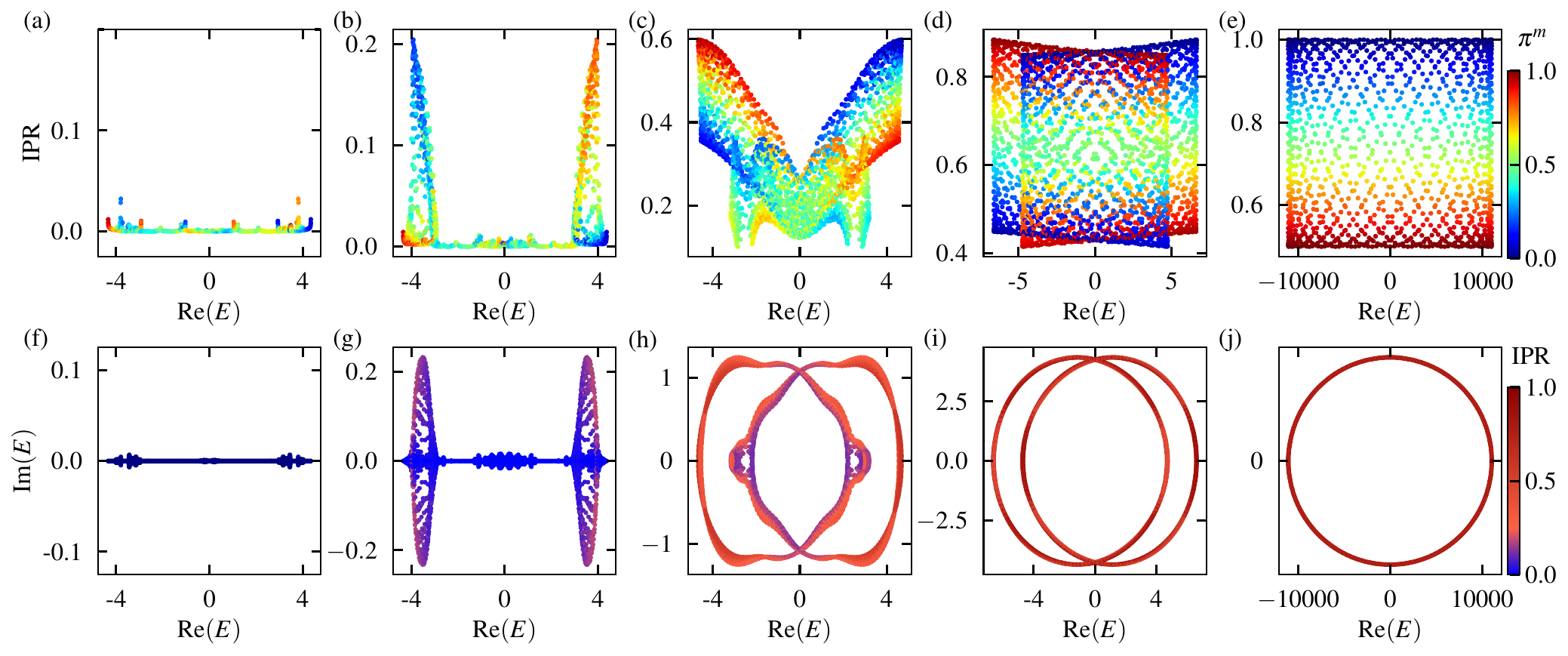}
    \caption{(a-e) The upper panel shows the inverse participation ratio (IPR) as a function of the real energy. Color-bar represents spin alignment ($\pi$), while the lower panels (f-j) display the corresponding spectra in the complex plane (Re($E$) versus Im($E$)), with color indicating the IPR. Results are presented for $\Gamma = 0.5, 1.0, 1.6, 2.3,$ and $ 10.0$. The other parameters are $\alpha_y = \alpha_z = 1.0$, $\lambda = 1.0$, $t=1.0$, $h_z=1.0$, and system size $N = 2584$.}
    \label{fig:3}
\end{figure*}

With the knowledge of localization transition for the system under consideration in hand, we move on to examine the properties of the states in the spectrum. To this end, we plot the IPR of individual states in different regions by varying $\Gamma$ for a cut through the phase diagram at a particular $h_z$. Interestingly, we find that for a range of values of $h_z$, when in the localized phase, the states in the spectrum together exhibit non-trivial localization properties. In Fig.~\ref{fig:3} we plot the IPR of each state with respect to their real eigen energies for different values of $\Gamma$ while keeping $h_z=1.0$. It can be seen that when in the delocalized phase ($\Gamma=0.5$), the IPR of all the states are vanishingly small as expected (Fig.~\ref{fig:3}(a)). When $\Gamma=1.0$, states with finite and vanishingly small IPR values appear, indicating the intermediate phase (Fig.~\ref{fig:3}(b)). Interestingly, however, when in the localized phase ($\Gamma=1.6$), the IPR of the states follows a butterfly-like distribution as shown in Fig.~\ref{fig:3}(c), and we call this the butterfly spectral map. This butterfly spectral map shows a non-trivial distribution of states in terms of their extent of localization.

To understand the origin of this peculiar nature of the spectrum, we resolve the localization properties of individual spin components. In Fig.~\ref{fig:2}(c) and (d), we depict the IPR for both components with respect to their real energy eigenvalues. It can be seen that, while a pair of wings-like features appears for both the components, the maps include states with vanishingly small IPR. However, when combined, the butterfly spectral map appears, which is the true feature of the spectrum, ranging from vanishing small to finite values. Although the individual components show a wider distribution of the IPR near the band edges compared to the center, when combined, they generate the butterfly spectral map depicting the true physical nature of the spectrum.  

It is to be noted that the butterfly spectral map is sensitive to the value of $\Gamma$. With an increase in $\Gamma$, the butterfly nature gradually disappears as shown in Fig.~\ref{fig:3}(d). In the limit of large $\Gamma$, the onsite potential strength becomes stronger, and one would expect all the states exhibiting a similar extent of localization or similar IPR values. Surprisingly, in this case, we obtain that with an increase in $\Gamma$, the spread of IPR values of the states or the spectral map becomes much wider. Moreover, a pair of states come close to each other in energies and also in their IPR values, and eventually, they become close to degenerate in their IPR values. This also results in a spectral map which resembles a caterpillar for very large values of $\Gamma$ as shown in Fig.\ref{fig:3}(e). Such resilience of the spectral map against $\Gamma$ can be attributed to the effect of $h_z$. 

In order to know the combined effect of textured Zeeman field and Rashba spin–orbit effects, we compute the local spin-projection operator at site $i$ as
\[
\Pi_i
      = \tfrac{1}{2}
      \begin{pmatrix}
        1 + \cos\theta_i & \sin\theta_i \\
        \sin\theta_i & 1 - \cos\theta_i
      \end{pmatrix}.
\]
Here $\theta_{i}$ is the polar angle, which encodes the spatially varying Zeeman
texture (for example, $\theta_{i}=2i\pi/(N)$ for an $N$-site ring).
This operator projects an arbitrary spinor onto the local
spin orientation $\hat{n}_{i},$
where $\hat{n}_i = (\sin\theta_i, 0, \cos\theta_i)$ specifies the local spin quantization direction. 
The expectation value $\psi_i^{\dagger}\Pi_i\psi_i \in [0,1]$ gives the probability of finding the spin at site $i$ aligned with $\hat{n}_i$. 
When $\theta_i = 0$, the local spin points entirely along the $+z$ direction and $\Pi_i$ projects onto the pure spin--up state with $\psi_i^{\dagger}\Pi_i\psi_i = 1$. 
Conversely, for $\theta_i = \pi$, the spin points along the $-z$ direction, corresponding to a fully anti--aligned spin (spin--down) with $\psi_i^{\dagger}\Pi_i\psi_i = 0$. 
For intermediate values of $\theta_i$, $\psi_i^{\dagger}\Pi_i\psi_i$ lies between $0$ and $1$, representing a partial alignment of the spin with the local Zeeman texture.

Given an eigenstate $\ket{\psi_{m}}$ of the non-Hermitian Hamiltonian, the
 local density projected by texture at site $i$ for the eigenstate $m$ is defined
as
\[
\pi^m = \bra{\psi_{m}} \big( \ket{i}\bra{i}\otimes \Pi_{i} \big) \ket{\psi_{m}}.
\]
Here $\ket{i}$ denotes the orbital part of the basis localized at site $i$,
and the tensor product structure ensures that the projection acts in the
spin subspace.

The quantity $\pi^{m}$ represents the projected local spin density of the $m$-th eigenstate at site $i$. 
It measures the probability of finding the particle at site $i$ with its spin aligned along (or opposite to) the local quantization axis defined by $\Pi_{i}$. 
\begin{itemize}
  \item $\pi^m \to 1$ indicates that the state $\ket{\psi_{m}}$ has a complete spin alignment at site $i$.
  \item $\pi^{m} \to 0$ corresponds to vanishing local spin density, i.e., negligible probability of finding the particle with that spin orientation at site $i$.
\end{itemize}
We embed the value of $\pi^m$ for each state in Fig.~\ref{fig:3} shown in colors. This shows a clear asymmetry in spin alignment with respect to the wings of the butterfly. Such asymmetry clearly depicts the effect of $h_z$ on the system in introducing spin alignment. 

\begin{figure}[t]
    \centering
    \includegraphics[width=1.0\linewidth]{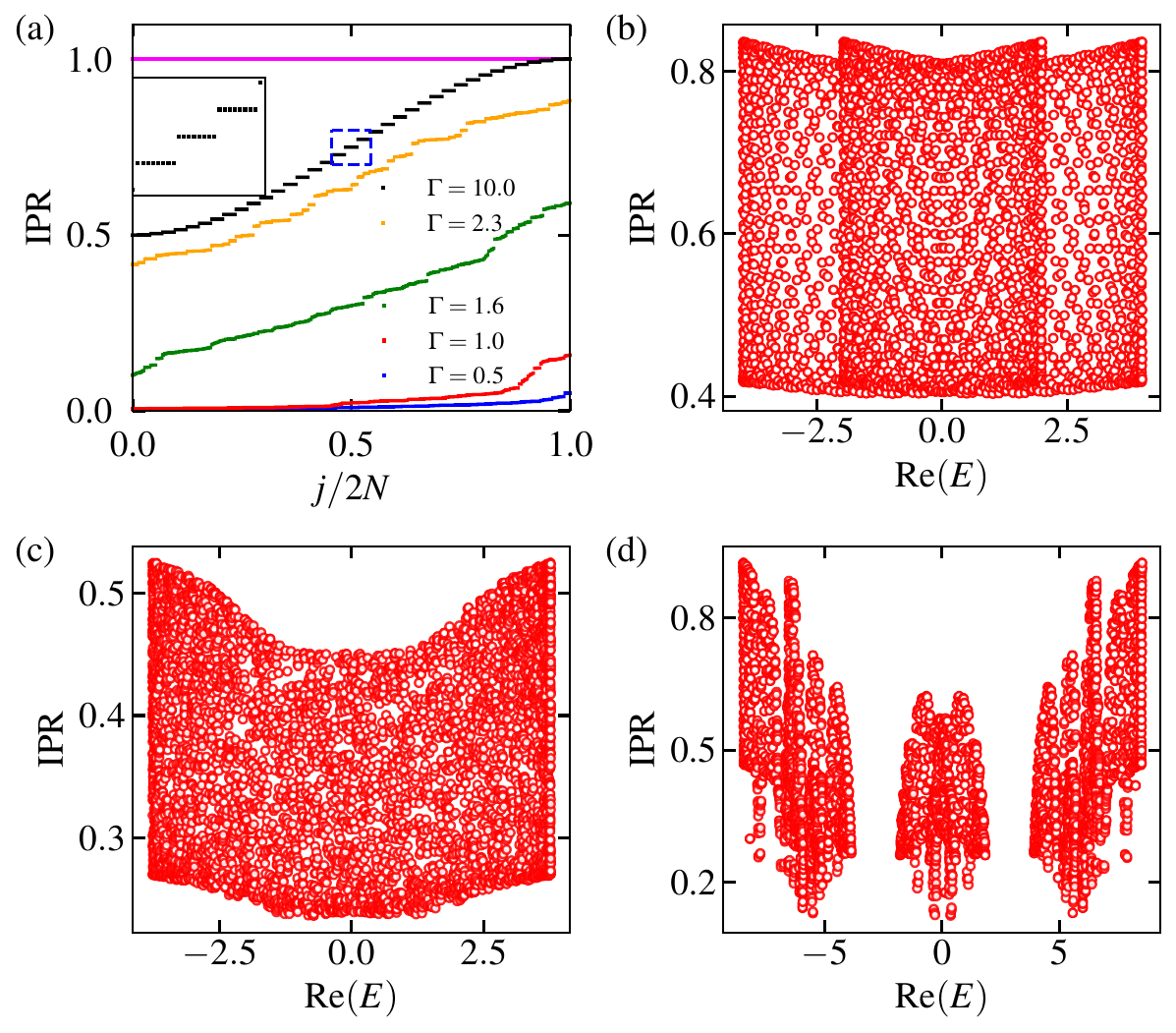}
    \caption{(a) IPR with corresponding state index $(j/2N)$ for different $\Gamma=0.5, 1.0, 1.6, 2.3, 10.0$ shown in blue, red, green, orange and black, respectively. The top magenta curve is also for $\Gamma=10$ but in the absence of RSO and $h_z$ terms. 
    The absence of butterfly spectra map shown for non-Hermitian limit in (b) $\alpha_z=\alpha_y=0.0, h_z=1.0$ and (c) $\alpha_z=\alpha_y=1.0, h_z=0.0$ while keeping $\Gamma=1.6$ and $\lambda=1.0$. (d) shows the Hermitian limit ($\Gamma=0.0$) with $\alpha_z=\alpha_y=h_z=1.0$. Here, the system size is $N=2584$ except for (a) where $N=144$ is considered for clarity.}
    \label{fig:4}
\end{figure}

The butterfly spectral map and its evolution also exhibits distinctly different signatures in the complex plane. In Fig.~\ref{fig:3} lower panel, we plot the energies in the complex plane and their associated IPR values (color-coded). In the case of a butterfly map, two intersecting loops appear (Fig.~\ref{fig:3}(i)). With an increase in $\Gamma$, the two loops tend to become circular and then eventually fall on top of each other. This also clearly suggests the degeneracy in IPR of the state at higher values of $\Gamma$. To understand this, we plot the IPR for all the states arranged according to their IPR (minimum to maximum) for different values of $\Gamma$ in Fig.~\ref{fig:4}(a). Here, $j/2N$ is the normalized state index. Clearly, for the delocalized phase ($\Gamma=0.5$), the IPR for all the states vanish. However, with increase in $\Gamma$, the intermediate phase appears where a portion of states exhibit finite IPR (see for $\Gamma=1.0$). In the localized phases ($\Gamma=1.6,~2.3,~10$), the IPR for all the states are finite. It is important to note that of the three curves for the localized states, the one with $\Gamma=1.6$ corresponds to the butterfly spectral map. For $\Gamma=10.0$ the states become degenerate in IPR with certain degree of degeneracy (see inset of Fig.~\ref{fig:4}(a)). Moreover, the entire curve follows a smooth sinusoidal pattern as a function of $j/N$ in this case unlike the one corresponding to the butterfly spectral map ($\Gamma=1.6$). This overall distribution of IPR of the states is a result of the RSO coupling and the magnetic field when the states become localized due to $\Gamma$. However, in the absence of both RSO coupling and textured magnetic field, all the state exhibit almost equal IPR $\sim 1$ (top most magenta curve in Fig.~\ref{fig:4}(a))

Now we will show that the butterfly spectral map is a combined effect of RSO coupling, textured magnetic field and the non-Hermitian quasiperiodic potential. By switching off any of these will lead to the absence of the butterfly spectral map. To show this we plot the IPR for all the real energies of the states for two situations: (i) $\alpha_y=\alpha_z=0$, $h_z=1$, (ii) $\alpha_y=\alpha_z=1$, $h_z=0$ in Fig.~\ref{fig:4} (b) and (c), respectively. Although both the spectral maps show a wide distribution of IPR, they don't exhibit any butterfly type pattern. We also examine the Hermitian limit by considering $\alpha_y=\alpha_z=1$, $h_z=1$ and $\Gamma=0$. In this limit, we keep a strong $\lambda$ to be in the localized phase to achieve finite IPR for the states which is shown in Fig.~\ref{fig:4}(d). In this case also we don't see any butterfly pattern in the spectral map due to the presence of the gaps in the spectrum.

\paragraph*{Conclusion.-} 
In summary, we have shown that a spinful non-Hermitian AAH model incorporating Rashba spin–orbit coupling and a spatially textured Zeeman field exhibits an emergence of a novel butterfly spectral map in the localized regime. This butterfly structure originates from the unusual pattern in the degree of localization of the states in the spectrum. Our analysis shows that the butterfly spectral map is not a property of only the quasiperiodic system, but instead it arises from the interplay of non-Hermitian quasiperiodic potential, spin–orbit induced spin mixing, and the textured magnetic field. We find that the Zeeman texture plays a dual role: it stabilizes an intermediate phase that hosts both localized and extended states and it simultaneously induces a pronounced asymmetric spin alignment across the wings of the butterfly. Resolving the spectrum into individual spin components reveals that the butterfly map is a genuinely spin-entangled spectral feature, which only becomes apparent when both spin sectors are taken together. Counterintuitively, at large non-Hermitian potential strengths, the system avoids uniform strong localization and instead exhibits a collapse of the butterfly into a caterpillar-like spectral structure, reflecting a subtle competition between non-Hermiticity and the textured Zeeman field.

\paragraph*{Acknowledgment.-}

T.M. acknowledges support from Science and Engineering Research Board (SERB), Govt. of India, through project No. MTR/2022/000382 and STR/2022/000023. D.C. acknowledges financial support from DST (project number DST/WISE-PDF/PM-40/2023).

\bibliography{ref}

\end{document}